\documentclass[pra,showpacs,twocolumn]{revtex4}

\newcommand{\be}{\begin{equation}}
\newcommand{\ee}{\end{equation}}

\usepackage{amsfonts}
\usepackage{amsmath}
\usepackage{graphicx}
\usepackage{amssymb}
\usepackage{amsmath}
\usepackage{amssymb}
\usepackage{graphicx}
\usepackage{lscape}
\usepackage{color}
\usepackage{epstopdf}

\begin{document}
\title{Universality and tails of long range interactions in one dimension}

\author{Manuel Valiente}
\affiliation{SUPA, Institute of Photonics and Quantum Sciences, Heriot-Watt University, Edinburgh EH14 4AS, United Kingdom}
\author{Patrik \"Ohberg}
\affiliation{SUPA, Institute of Photonics and Quantum Sciences, Heriot-Watt University, Edinburgh EH14 4AS, United Kingdom}

\begin{abstract}
Long-range interactions and, in particular, two-body potentials with power-law long-distance tails are ubiquitous in nature. For two bosons or fermions in one spatial dimension, the latter case being formally equivalent to three-dimensional $s$-wave scattering, we show how generic asymptotic interaction tails can be accounted for in the long-distance limit of scattering wave functions. This is made possible by introducing a generalisation of the collisional phase shifts to include space dependence. We show that this distance dependence is universal, in that it does not depend on short-distance details of the interaction. The energy dependence is also universal, and is fully determined by the asymptotic tails of the two-body potential. As an important application of our findings, we describe how to eliminate finite-size effects with long-range potentials in the calculation of scattering phase shifts from exact diagonalisation. We show that even with moderately small system sizes it is possible to accurately extract phase shifts that would otherwise be plagued with finite-size errors. We also consider multi-channel scattering, focusing on the estimation of open channel asymptotic interaction strengths via finite-size analysis.
 \end{abstract}
\pacs{
}
\maketitle

\section{Introduction}
Few-body systems play a central role in quantum mechanics, which arguably started with Schr{\"o}dinger's solution to the hydrogen atom \cite{Schrodinger}. Traditionally, the scattering states of a two-body system or, more specifically, their asymptotic form, are used to predict cross sections in atomic, molecular, nuclear and particle physics \cite{Taylor,Joachain}, which give useful information about the underlying interactions and structure of the colliding bodies. When the interactions have a short range, there is a technique that has been championed by nuclear physicists to extract effective nucleon-nucleon and multi-nucleon interactions using few-body observables only. This is the Effective Field Theory (EFT) of nuclear forces \cite{Machleidt,Epelbaum,vanKolck,Hammer,Epelbaum2,Phillips,Kaplan1,Kaplan2}, which began with Weinberg's seminal papers \cite{Weinberg1,Weinberg2}. In its modern form, especially when the system is discretised on a lattice, it is combined with energetic methods, i.e. exact diagonalisation or imaginary time evolution, rather than traditional scattering theory, to predict scattering observables (for a pedagogical review, see \cite{Nicholson}). The methodology involved is a generalisation of the ground-breaking work of L{\"u}scher \cite{Luescher}, who was the first to show the relationship between low-energy scattering observables and the two-body spectrum in a periodic box, at least in the weak-coupling limit. The advantages of this kind of approach are most obvious when at least one of the particles in the system is a composite object, since it is generally easier to extract the ground-state energy of a three- or four-body system in a finite volume than to solve the corresponding multichannel Faddeev or Yakubovsky equations \cite{Lee}, or in generic multi-channel problems.

The use of EFT is not confined to nuclear physics. In ultracold atoms, the lowest-order EFT, corresponding to the Huang-Yang pseudopotential \cite{HuangYang}, is routinely used in the theory of Bose-Einstein condensates \cite{Dalfovo} and spin-$1/2$ Fermi gases \cite{Giorgini}. It plays a particularly important role where it is most accurate, that is, near an $s$-wave two-body resonance in a two-component Fermi gas, for which a set of universal relations hold in all spatial dimensions \cite{Tan1,Tan2,Tan3,ValienteZinnerMolmer,ValienteZinnerMolmer2,WernerCastin}. Effects beyond lowest-order EFT, including three-body effects, are patent in the three-boson problem, where these play a major role in Efimov physics \cite{Bedaque}, observed for the first time with ultracold atoms by Kraemer {\it et al.} \cite{Grimm}. The development of EFTs in reduced (one and two) spatial dimensions is somewhat behind that of the three-dimensional case. There are two well-known examples of lowest-order EFT in one dimension (1D), namely the Dirac-delta interaction for bosons and spin-$1/2$ fermions which is UV-regular, and its odd-wave  dual (sometimes termed "$p$-wave"), which is UV-divergent but renormalisable, as first discussed by Cheon and Shigehara \cite{CheonShigehara} in the position representation, and later on in the momentum representation \cite{ValienteZinnerEFT,CuiEFT}. The EFT to next-to-leading-order for two-body scattering in 1D was also discussed in ref. \cite{ValienteZinnerEFT}. 

The effective interactions discussed above are all concerned with low-energy scattering. In one spatial dimension, however, we have shown \cite{ValienteOhberg} that Luttinger liquids whose constituents are scalar particles may depend very little on low-energy interactions except for extremely low densities, since the relevant energy scale for two-particle collisions is twice the Fermi energy. When realistic interactions are involved, such as Born-Oppenheimer potentials, which have Van der Waals tails, and especially in multichannel problems, it can be considerably easier to numerically diagonalise the two-body Hamiltonian in a finite box rather than solving the Schr{\"o}dinger or Lippmann-Schwinger equations in infinite space. The scattering phase shifts can then, in principle, be extracted from extensions to L{\"u}scher's analysis, for which there exists a vast and comprehensive literature \cite{BeaneTwoNucleons,Fukugita,TanBose,ValienteZinnerLuscher,PRD1,PRA1}. Unfortunately, these approaches typically \footnote{An exception is ref. \cite{BeaneCoulomb}, which deals with emergent, periodised Coulomb interactions relevant in Nuclear Physics.} assume that the interaction tails are negligible, a perfectly valid assumption in the context of these works. This, however, is hardly the case in atomic and molecular physics: interactions between two neutral atoms, a neutral atom and an ion, and two dipolar molecules, display $r^{-6}$, $r^{-4}$ and $r^{-3}$ tails, respectively. Therefore, exact diagonalisation in a finite box can introduce very significant errors if the tails are not properly accounted for in the finite-size analysis of the phase shifts. 

Here we study scattering in one dimension which, for the fermionic case, is formally equivalent to three-dimensional $s$-wave collisions, for interactions that exhibit a long-range tail. Firstly, we show how the introduction of space-dependent phase shifts yields universal, energy-independent information relating the long-range tail and the spatially varying part of the phase shift. We then show how to obtain scattering phase shifts using finite-size energy considerations only, by taking into account this universal asymptotic behaviour. We also generalise our results to multi-channel scattering.

\section{Hamiltonian of the system} 
We consider the non-relativistic two-body scattering problem whose dynamics is governed by the following position-represented Hamiltonian 
\begin{equation}
\mathcal{H}=\frac{p_1^2}{2m}+\frac{p_2^2}{2m}+V(x_1-x_2).\label{Ham}
\end{equation}
Above, $m$ is the mass of the particles, while $V$ is a generic interaction potential, assumed to have a vanishing asymptotic limit, i.e. $\lim_{x\to \pm \infty} V(x)=0$. The stationary Schr{\"o}dinger equation $\mathcal{H}\Psi=\mathcal{E}\Psi$ is separable in terms of centre of mass $X=(x_1+x_2)/2$ and relative $x=x_1-x_2$ coordinates in the usual way, i.e. 
\begin{equation}
\Psi(X,x)=e^{iKX}\psi(x),
\end{equation}
where $K=k_1+k_2$ is the total momentum of the system, while $\psi(x)$ is the relative wave function. This satisfies the stationary Schr{\"o}dinger equation $H\psi=E\psi$, where $E=\mathcal{E}-\hbar^2K^2/4m$, and $H$ is given by
\begin{equation}
H=\frac{p^2}{2\mu}+V(x),\label{Hamrel}
\end{equation}
with $\mu=m/2$ the reduced mass, and $p$ the relative momentum operator. Positive relative energies $E=\hbar^2k^2/2\mu$ correspond to scattering states, which we shall focus on in the following.

\section{General framework} 
In order to establish the universal and non-universal properties of two-body collisions of identical particles, we consider the asymptotic, or long distance limit of the Schr{\"o}dinger equation $H\psi=E\psi$. Since $V(x)\to 0$ at long distances, the asymptotic form of the stationary scattering states, $\psi^{\mathrm{asymp}}_k$, must have the form
\begin{align}  
\psi^{\mathrm{asymp}}_{\mathrm{B}}(x)&=\sin(k|x|+\theta_k^{\mathrm{B}}(x)),\label{asympbosons}\\
\psi^{\mathrm{asymp}}_{\mathrm{F}}(x)&=\mathrm{sgn}(x)\sin(k|x|+\theta_k^{\mathrm{F}}(x))\label{asympfermions}.
\end{align}
Above, the subscript/superscript B (F) refers to bosons or spin-singlet fermions (spin-triplet or spinless fermions). Notice that we have given the phase shifts $\theta^{\mathrm{B}}$ and $\theta^{\mathrm{F}}$ explicit space dependence. This is crucial for the analysis that follows. We now replace the interaction potential by its asymptotic form $V^{\infty}(x)$, and insert the asymptotic scattering states, Eq.~(\ref{asympbosons}) or (\ref{asympfermions}) into the Schr{\"o}dinger equation. In this way, the following asymptotic differential equation is obtained for the position-dependent phase shift (we drop the superscript $B$ or $F$ for ease of notation)
\begin{equation}
\frac{1}{2}\left(\frac{d\theta_k}{dx}(x)\right)^2+k\mathrm{sgn}(x)\frac{d\theta_k}{dx}(x)+\frac{\mu V^{\infty}(x)}{\hbar^2}=0,\label{diffeq1}
\end{equation}
subject to the conditions that $|\theta_k''(x)/V^{\infty}(x)|\to 0$ and $|\theta_k''(x)/\theta_k'(x)|\to 0$. If, moreover, $\theta_k''(x)$ is much smaller (in absolute value) than $(\theta_k'(x))^2$, as is usually the case (see below for power-law potentials), then the equation that must be solved for the sake of consistency is even simpler,
\begin{equation}
k\mathrm{sgn}(x)\frac{d\theta_k}{dx}(x)+\frac{\mu V^{\infty}(x)}{\hbar^2}=0,
\end{equation}
which implies
\begin{equation}
\theta_k(x)=\theta_k-\frac{\mu}{\hbar^2k}\int^x dx \mathrm{sgn}(x)V^{\infty}(x).\label{simpletheta}
\end{equation}
As a convention, we choose the integration constant in Eq.~(\ref{simpletheta}) such that $\theta_k=\lim_{x\to \infty} \theta_k(x)$ corresponds to the scattering phase shift.

\section{Asymptotic power-law interactions} 
The most physically relevant interactions in nature typically exhibit long-distance power-law tails. Such is the case of Coulomb, dipolar or van der Waals tails that dictate electron-electron or atom-atom interactions. 

Power-law interaction tails are described by the asymptotic potential
\begin{equation}
V^{\infty}(x)=\frac{g_{\nu}}{|x|^{\nu}}.
\end{equation}
The strength of the tails can be arbitrarily large and both repulsive ($g_{\nu}>0$) or attractive ($g_{\nu}<0$). The space-dependent phase shift must be calculated in this case from Eq.~(\ref{simpletheta}) and is given by
\begin{align}
\theta_k(x)&=\theta_k+\frac{\mu}{\hbar^2 k}\frac{g_{\nu}}{\nu-1}\frac{1}{|x|^{\nu-1}},\hspace{0.1cm} \nu \ne 1 ,\label{nugt1}\\
\theta_k(x)&=\theta_k-\frac{\mu}{\hbar^2 k}g_1\log|2kx|,\hspace{0.1cm} \nu=1.\label{logarithmicphaseshift}
\end{align}
The above results are remarkable. While the constant part of the phase shifts $\theta_k$ must be calculated microscopically, the space-varying asymptotes are fully universal: not only are their functional forms fixed by the asymptotic tails of the interaction, but also their pre-factors are determined by these. In particular, the famous logarithmic phase shifts of the Coulomb potential \cite{Abramowitz} appear naturally within the current framework, and the only non-universal feature of the phase shift is the constant $\theta_k$. Notice, once more, that the microscopic short-distance details of the interactions do not enter the asymptotes besides in the value of $\theta_k$, and smoothened interactions at short distances play no role here. 

\section{Calculating the constant part of the phase shift} 
\begin{figure}[t]
\includegraphics[width=0.5\textwidth]{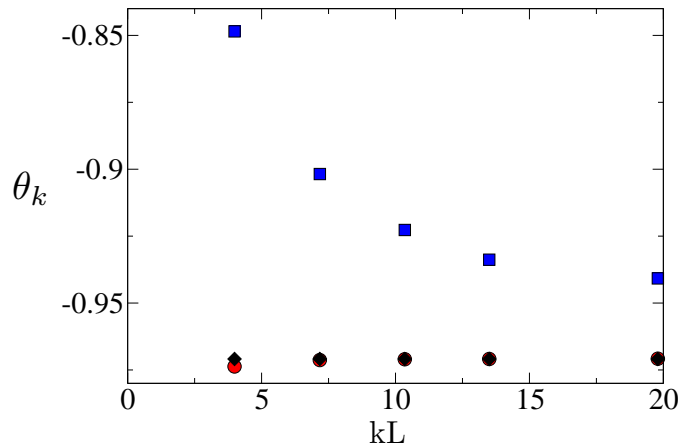}
\caption{Calculated phase shifts for the Calogero-Sutherland model with $2\mu g_2/\hbar^2=1$. Calculations using Eq.~(\ref{thetak}) with (red circles) and without (blue squares) space-dependent phase shifts are compared to the exact phase shift (black diamonds).}
\label{fig:theta-calogero}
\end{figure}

The above analysis is not only useful from a theoretical point of view, but can also be used to extract numerically, in a very straightforward manner, the microscopic scattering phase shifts $\theta_k$ for interactions falling off faster than $1/|x|$, and the so-called Coulomb phase shifts for Coulomb interactions. To see this, consider the following fictitious problem: place a single particle of mass $\mu$ in a finite box of size $2L$ with open boundary conditions, i.e. $\psi(L)=\psi(-L)=0$, with the full (i.e. not just the asymptotic part) interaction potential $V(x)$ centered at $x=0$. Diagonalise the single-particle problem numerically and extract the eigenvalues. For sufficiently large $L$, such that the asymptotic potential is accurate for $x\sim L$, the eigenfunctions have the asymptotic behaviour in Eqs.~(\ref{asympbosons}) or (\ref{asympfermions}) for bosons and fermions, respectively. The energy of the eigenstates is given by $\hbar^2k^2/2\mu$, and $k$ is quantised by applying open boundary conditions to the asymptotic eigenfunctions (\ref{asympbosons}) or (\ref{asympfermions}). The quantisation reads
\begin{equation}
k=\frac{\pi n}{L}-\frac{\theta_k(L)}{L}, \hspace{0.1cm} n\in \mathbb{Z}_+\label{quantisationcondition}
\end{equation}
Since the eigenvalues are known after numerical diagonalisation, so is $k=\sqrt{2\mu E/\hbar^2}$. By rewriting the space-dependent phase shift as $\theta_k(x)=\theta_k+\Delta_k(x)$, where $\Delta_k(x)$ is the universal space-dependent part of the phase shift, Eq.~(\ref{quantisationcondition}) is already solved and gives
\begin{equation}
\theta_k=\pi n -kL-\Delta_k(L).\label{thetak}
\end{equation}
There is a particularly simple model, namely the Calogero-Sutherland model \cite{Sutherland}, for which it is possible to obtain all eigenvalues given one of them. In this model, which has inverse square interactions, the phase shift $\theta^{(0)}_k=-\pi(\lambda-1)/2$ \cite{Sutherland}, with $g_2=(\hbar^2/m)\lambda(\lambda-1)$ is a constant (up to a $\mathrm{sgn}(k)$ factor), due to the fact that the model in free space is scale invariant. Scale invariance is only broken by the existence of a boundary in the system. In Fig.~\ref{fig:theta-calogero} we show the scattering phase shifts calculated numerically by diagonalising the system (see Appendix \ref{Appendix1}) and either including or neglecting the universal space dependence of the phase shift. There, it is clear that the space dependence is absolutely necessary at low energies. The smaller the $kL$ value is the larger the deviation is between the exact value and the result using the position-dependent phase shift. However, it is worth stressing that at $kL=3.99005$ (the lowest possible value of $kL$) the error is still a mere $0.28\%$. 

%%%%%%%%%%%%%%%%%%%%%%%%%%%%%
%%%%%%%%%%%%%%%%%%%%%%%%%%%%%%

\begin{figure}[t]
\includegraphics[width=0.5\textwidth]{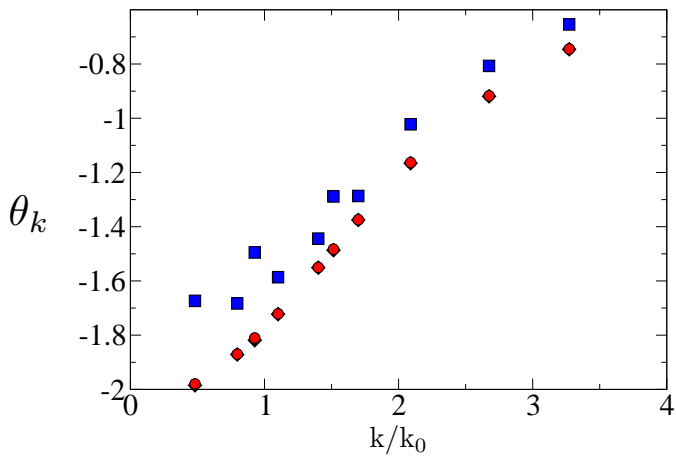}
\caption{Calculated fermionic phase shifts with the interaction in Eq.~(\ref{k0g2}), with $2\mu g_2/\hbar^2=3$. Calculations using Eq.~(\ref{thetak}) with (red circles) and without (blue squares) space-dependent phase shifts, compared to the phase shift from integration of the Lippmann-Schwinger equation (black diamonds). Data are obtained for two different sizes, $L_1$ and $L_2$, and $k_0L_1=1$. Positive values of $\theta_k$ have been shifted by $-\pi$ for ease of visualisation.}
\label{fig:theta-defect-calo}
\end{figure}

\begin{figure}[t]
\includegraphics[width=0.5\textwidth]{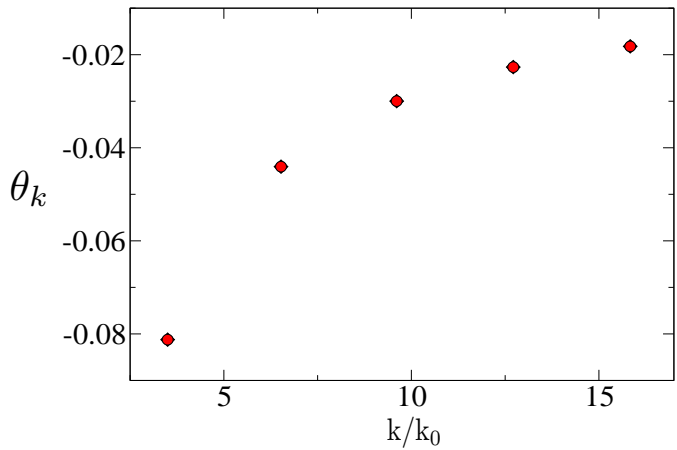}
\caption{Calculated phase shifts for the Coulomb interaction with $2\mu g_1L /\hbar^2=1$ and $k_0L=1$. Calculations using Eq.~(\ref{thetak}) with space-dependent phase shifts (red circles) are compared to the exact phase shifts (black diamonds) obtained from Coulomb functions (see text).}
\label{fig:theta-Coulomb}
\end{figure}

The pure Calogero-Sutherland interaction, being scale invariant and exactly solvable, is however not the best example to illustrate the power of the method. We can also use a softened version of the $1/x^2$ potential, given by
\begin{equation}
V(x)=\frac{k_0^2g_2}{1+(k_0x)^2},\label{k0g2}
\end{equation}
for which the asymptotic potential is also $V^{\infty}(x)=g_2/x^2$. In Fig.~\ref{fig:theta-defect-calo}, the results for the fermionic phase shift from exact diagonalisation with and without $\Delta_k$ are compared with well-converged values obtained from a numerical solution to the Lippmann-Schwinger equation (see Appendix \ref{Appendix2}). Again, the highest error is small, approximately a $0.5\%$. As a last application, in Fig.~\ref{fig:theta-Coulomb} we show the constant part of the Coulomb phase shift, calculated from exact diagonalisation, and compared to the exact result \cite{Abramowitz} $\theta_k=\mathrm{Arg}\hspace{0.1cm}\Gamma(1+i\mu g_1/\hbar^2k)$.

%%%%%%%%%%%%%%%%%%%%%%%%%%%%%
%%%%%%%%%%%%%%%%%%%%%%%%%%%%%%

\section{Tail renormalisation in multi-channel collisions}\label{multichannel}
So far, we have discussed the standard single-channel scattering with long-range interactions. Single-channel problems, however, can be quite efficiently and accurately solved by means of a brute force numerical implementation of the Lippmann-Schwinger equation. The solution to this yields the corresponding phase shifts and, complemented with the universal asymptotic spatially-dependent "phase shifts" explained in the previous section, all relevant information about the scattering asymptotes can be extracted. Multi-channel problems, however, are significantly more difficult to solve via brute force integration of the Lippmann-Schwinger equation. In general, the incident waves of each channel must be carefully prepared before attempting a numerical solution. Moreover, if the multichannel nature of the system is due to multi particle bound states, which effectively describe single particles, preparing the incident state and solving the integral equations numerically can prove to be quite challenging. An easy-to-visualise example would be the elastic collision of a two-body bound state with a heavy (static) particle. Notice that, even if all bare interactions in a given system have the same asymptotic tails, the strength of the tails of the open channel interactions, and in some cases also their functional form \footnote{A well known example is hydrogen-hydrogen (atom) elastic collisions. All particles involved interact via Coulomb potentials, but the Born-Oppenheimer interaction has van der Waals tails.}, are modified by the inclusion of all channels in the system. It is therefore of great interest to develop methodology that not only delivers scattering phase shifts, but also gives information about the tails of the effective open channel interaction. 

The general theory of multi-channel asymptotes is, fortunately, not very different from the single-channel picture. We denote the different physical channels by $\alpha_i$, and assume that the open channel ($\alpha_1$) where elastic collisions are to be investigated has the lowest energy in its non-interacting ground state. The asymptotic form of the effective interaction in the elastic channel $(\alpha_1,\alpha_1)$ is denoted by $V^{\infty}_{\alpha_1,\alpha_1}(x)$, such that the asymptotic scattering state in the open channel, $\psi_{\alpha_1}(x)$, satisfies
\begin{equation}
-\frac{\hbar^2}{2\mu}\frac{\partial^2\psi_{\alpha_1}(x)}{\partial x^2}+V^{\infty}_{\alpha_1,\alpha_1}(x)\psi_{\alpha_1}(x)=(E-\mathcal{E}_{\alpha_1})\psi_{\alpha_1}(x),
\end{equation}
where $\mathcal{E}_{\alpha_1}$ is the non-interacting ground state energy, on the infinite line, of the two-body elastic channel $(\alpha_1,\alpha_1)$.

The most physically relevant situation corresponds to a power-law tail $V_{\alpha_1,\alpha_1}^{\infty}=g_{\nu}^{\alpha_1}/|x|^{\nu}$. This is the case, for instance, in multichannel models of Feshbach resonances \cite{Chin}. If only a finite number of channels are present, then the power $\nu$ and strength $g_{\nu}^{\alpha_1}$ of the tails can be easily calculated, as in the example shown below. When this is the case, the phase shifts can be calculated just as in single-channel scattering. If, on the other hand, there is an infinite number of channels (e.g. in models of confinement-induced resonances \cite{Olshanii,Saenz1,Saenz2,Saenz3,ValienteCIR,Schmelcher}), the exponent $\nu$ may change, and the strength must be calculated non-perturbatively. It is therefore important, especially for infinitely many channels, to devise a way to extract the strength of the tail in the open channel by energetic arguments only. To do this, take two boxes of lengths $2L_1$ and $2L_2$ ($L_1\ne L_2$)  both having one eigenstate corresponding to a certain momentum $k$, linked with the integers $n_1$ and $n_2$ ($n_1\ne n_2$), respectively (see Eq.~(\ref{quantisationcondition})). Since both states have the same momentum, their phase shifts $\theta_k$ are identical. By equating the corresponding conditions for $\theta_k$ with lengths $2L_1$ and $2L_2$ in Eq.~(\ref{thetak}), and using that $\Delta_k(x)=\mu g_{\nu}^{\alpha_1}/[\hbar^2 k(\nu-1)|x|^{\nu-1}]$, the following is found for the strength of the tail
\begin{equation}
g_{\nu}^{\alpha_1}=(\nu-1)\frac{\hbar^2k}{\mu}\frac{(L_1L_2)^{\nu-1}}{L_1^{\nu-1}-L_2^{\nu-1}}\left[\pi(n_2-n_1)+k\left(L_1-L_2\right)\right].\label{equationmulti}
\end{equation} 

%%%%%%%%%%%%%%%%%%%%%%%%%%%%%
%%%%%%%%%%%%%%%%%%%%%%%%%%%%%%

To illustrate and benchmark the above results in multichannel calculations, we consider a simple two-channel model, which is similar to what is sometimes used in modelling magnetic Feshbach resonances \cite{Chin}. The Schr\"odinger equation has the form
\begin{align}
-\frac{\hbar^2}{2\mu}\frac{\partial^2\phi_1(x)}{\partial x^2}+V_1(x)\phi_1(x) -J\phi_2(x) &= E\phi_1(x),\label{bare1}\\
-\frac{\hbar^2}{2\mu}\frac{\partial^2\phi_2(x)}{\partial x^2}+V_2(x)\phi_2(x) -J\phi_1(x) &= E\phi_2(x).\label{bare2}
\end{align}
Above, $J>0$ is a constant that represents the coupling between the bare channel wave functions $\phi_1$ and $\phi_2$. The bare channels, $1$ and $2$, must be re-expressed in terms of physical channels, $A$ and $B$, by means of the transformation $\psi_A=\phi_1+\phi_2$ and $\psi_B=\phi_1-\phi_2$, up to a global normalisation constant. Defining $V_A(x)=V_B(x)=(V_1(x)+V_2(x))/2$ and $J_A(x)=J_B(x)=(V_1(x)-V_2(x))/2$, the multichannel equations become
\begin{align}
-\frac{\hbar^2}{2\mu}&\frac{\partial^2\psi_A(x)}{\partial x^2}+V_A(x)\psi_A(x)\nonumber\\
&+J_A(x)\psi_B(x)
= (E-\mathcal{E}_A)\psi_A(x),\label{psiAeq}\\
-\frac{\hbar^2}{2\mu}&\frac{\partial^2\psi_B(x)}{\partial x^2}+V_A(x)\psi_B(x)\nonumber\\
&+J_A(x)\psi_A(x)= (E-\mathcal{E}_B)\psi_B(x),\label{psiBeq}
\end{align}
where $\mathcal{E}_A=-J$ ($\mathcal{E}_B=+J$) is the open (closed) channel's non-interacting ground state energy. In the numerical example, we use fermions and choose $\nu=3$ (dipolar-like tails), and $V_i(x)=k_0^3g_3^{(i)}/(1+(k_0x)^3)$, $i=1,2$, with $g_3^{(1)}/g_3^{(2)}=2$. From Eq.~(\ref{psiAeq}), we clearly have $g_3^{A}=3g_3^{(1)}/4$. In order to test how accurate the estimation of the tails in the open channel can be in more involved calculations (given finite-size effects and numerical uncertainty in "exact" diagonalisation), we set $L_1$ to a constant, for which its ground state momentum $k=0.3406k_0$ is extracted. Then, we numerically find three other values, $L_2$, $L_3$ and $L_4$ such that the first, second and third excited states, respectively, correspond to momentum $k$. Using these values in Eq.~(\ref{equationmulti}), we obtain 6 different estimates of $g_3^{A}$, giving the result $g_3^{A}/g_3^{(1)}=0.764\pm 0.028$ (see Appendix \ref{Appendix3}). The exact value of $3/4$ is reproduced within error bars. Higher accuracy can be obtained by using many different values of $L_1$, due to the energy-independence of the tail strengths in this case. The phase shift is then estimated using Eq.~(\ref{thetak}) to be $\theta_k=-0.2693\pm 9\cdot 10^{-4}$. 

\section{Conclusions}
We have studied the asymptotic form of scattering states with interactions that have a long-range tail in one dimension which, in the odd-wave channel, also describe three-dimensional $s$-wave scattering. We have shown how a generalisation of phase shifts in the asymptotic scattering states yields a universal space-dependence of these, independent of energy and short-distance details of the interaction. These results have been used to generalise widespread methods in Nuclear Physics to extract scattering information via finite-size analysis to the atomic situation where long-range tails are very important. In particular, we have exemplified our findings with different power-law tails, including the Coulomb interaction, and found that it is possible to extract phase shifts accurately using exact, finite-size diagonalisation, even for relatively small system sizes. We have studied generic multi-channel problems, for which the single-channel results are valid as well, and obtained an expression that relates finite-size information with the strength of the asymptotic tails of the interaction in the open elastic channel. We studied an example of a two-channel model showing that our method is a viable alternative to coupled-channel numerical calculations of scattering states. Our results and methods may be especially relevant for multi-channel problems with an infinite number of coupled channels, as is the case in dimensional reduction and confinement induced resonances. The methods we have introduced can also be of great use for the non-expert in few-body physics, as it is generally easier to implement than other approaches.

\appendix
\section{Exact diagonalisation}\label{Appendix1}
We explain here the details of the exact diagonalisation used to obtain the results in the main text. We have chosen to discretise the system on a grid, since this is the simplest possible method, and is capable of giving accurate results if analysed properly.

We discretised the Laplacian using a third-order quadrature rule. We use $L_s$ equally spaced grid points with a lattice spacing $d$ such that the length of the box is $L_sd$. The stationary Schr{\"o}dinger equation is therefore discretised as
\begin{align}
-&\sum_{\mu=1}^{3}J_{|\mu|}\left[\psi(n+\mu)+\psi(n-\mu)\right]\nonumber \\
&+2(J_1+J_2+J_3+V(n)-E)\psi(n)=0,
\end{align}
where $J_{1}=(3/2d^2)\hbar^2/2\mu$, $J_{2}=-(3/20d^2)\hbar^2/2\mu$ and $J_3=(1/90d^2)\hbar^2/2\mu$. We diagonalise the Hamiltonian numerically for $L_s$ between $201$ and $801$, with $d$ adjusted so that $L$ is kept constant, and the energy of a given state as a function of the number of grid points is $E(L_s)$. We then fit the continuum limit $E_{\infty}$ as 
\begin{equation}
E(L_s)=E_{\infty}+\alpha L_s^{-1}+\beta L_s^{-2}.
\end{equation}
The error in the least square fits to $E_{\infty}$ are of the order of $10^{-6}\%$ in all cases we studied. The same types of fits are done for the phase shifts and the extracted values of the incident momentum of the scattering states, with similar fitting errors.

\section{Lippmann-Schwinger equation}\label{Appendix2}
The Lippmann Schwinger equation for bosons and fermions in one dimension reads (see, for instance \cite{ValientePhillipsZinnerOhberg})
\begin{equation}
\psi(x)=\psi_0(x) + \frac{\mu}{\hbar^2k}\int_{-\infty}^{\infty} dy \sin(k|x-y|)V(y)\psi(y),\label{LSE}
\end{equation}
where $\psi_0(x)=\cos(kx)$ for bosons and $\psi_0(x)=\sin(kx)$ for fermions. Notice that the Lippmann-Schwinger equation for one-dimensional fermions and three-dimensional $s$-wave scattering, after rearrangement of the integration limits in Eq.~(\ref{LSE}), are equivalent \cite{Joachain}. The phase shifts $\theta_k^{\mathrm{B}}$ and $\theta_k^{\mathrm{F}}$ for bosons and fermions, respectively, are obtained from the scattering wave functions as
\begin{align}
\cot \theta_k^{\mathrm{B}}&=\frac{\mu}{\hbar^2k}\int_{-\infty}^{\infty} dy V(y) \cos(ky)\psi_k(y), \label{thetaB}\\
\tan \theta_k^{\mathrm{F}}&=-\frac{\mu}{\hbar^2 k}\int_{-\infty}^{\infty}dy V(y) \sin(ky)\psi_k(y).\label{thetaF}
\end{align}

The numerical solution of Eq.~(\ref{LSE}) is obtained by discretising the integral in Eq.~(\ref{LSE}) using Gauss quadrature with a large distance cutoff $\Lambda$, and solving the resulting system of linear equations. The phase shifts, from Eqs.~(\ref{thetaB}) and (\ref{thetaF}), are obtained using Gauss quadrature.

\section{Details of the multi-channel calculation}\label{Appendix3}
The two interaction potentials $V_1$ and $V_2$ of the bare channels in Eqs.~(\ref{bare1}) and (\ref{bare2}) are given by
\begin{equation}
V_i(x)=\frac{k_0^3g_3^{(i)}}{1+(k_0x)^3}.
\end{equation}
In the example in the text, we chose $2\mu k_0g_3^{(1)}/\hbar^2=1$ and $g_3^{(2)}=g_3^{(1)}/2$. The length $L_1$ is chosen such that $k_0L_1=10$, which is large enough that the interaction potentials $V_i(x)$ are well in the asymptotic regime at $x=L_1$. The value of the inter channel coupling constant is $J=\hbar^2 k_0^2/2\mu$. The ground state momentum (i.e. corresponding to $n=1$ in Eq.~(\ref{quantisationcondition})) as obtained by numerically diagonalising the two-channel Hamiltonian in the bare $\{1,2\}$ basis is given in this case by $k/k_0=0.34060$. We then found three other lengths $L_i$, $i=2,3,4$ such that the $(i-1)$-th excited state corresponds to $k/k_0=0.34060$. These lengths were found to be $L_2=1.92362L_1$, $L_3=2.84622L_1$ and $L_4=3.76868L_1$. Using Eq.~(\ref{equationmulti}) for all pairs with different $(L_i,L_j)$, six different estimates for $g_{3}^{A}/g_3^{(1)}$ are found: $\{0.786361,0.794766,0.751591,0.742459,0.726599\}$. Their average $\langle g_3^{A}\rangle/g_3^{(1)}$ and standard deviation $\sigma$ are therefore given by
\begin{align}
\frac{\langle g_3^{A} \rangle}{g_3^{(1)}} &= 0.764,\\
\sigma&=0.028,
\end{align}
which are the values reported in Sec. \ref{multichannel}.

\acknowledgments
The authors acknowledge support from EPSRC grant No. EP/M024636/1. The work of MV was initiated at the Aspen Center for Physics, which is supported by National Science Foundation grant PHY-1066293. 

\bibliographystyle{unsrt}

\end{document}